\documentclass[aps,prd,preprint,eqsecnum,nofootinbib]{revtex4}

\usepackage{axodraw,amsmath,graphicx}

\def\bra#1#2{[#1\,#2]}
\def\ket#1#2{\langle #1\,#2\rangle}
\def\fig#1{Fig.~{\ref{#1}}}

\begin{document}

\noindent \hfill Brown-HET-1584

\vskip 1 cm

\title{The Tree Formula for MHV Graviton Amplitudes}

\author{Dung Nguyen}

\affiliation{Brown University, Providence, Rhode Island 02912, USA}

\author{Marcus Spradlin}

\affiliation{Brown University, Providence, Rhode Island 02912, USA}

\author{Anastasia Volovich}

\affiliation{Brown University, Providence, Rhode Island 02912, USA}

\author{Congkao Wen}

\affiliation{Brown University, Providence, Rhode Island 02912, USA}

\begin{abstract}
We present and prove a formula for the MHV scattering amplitude
of $n$ gravitons at tree level.  Some of the more interesting features of the
formula, which set it apart as being significantly different from
many more familiar formulas, include
the absence of any vestigial reference to a cyclic ordering of
the gravitons---making it in a sense a truly
gravitational formula, rather than a recycled Yang-Mills result,
and the fact that it simultaneously manifests both $S_{n-2}$ symmetry
as well as
large-$z$ behavior that is
${\cal O}(1/z^2)$ term-by-term, without relying on delicate cancellations.
The formula is
seemingly related to others by an enormous simplification provided by
${\cal O}(n^n)$ iterated Schouten identities, but our proof relies on a
complex analysis argument rather than such a brute force manipulation.
We find that the formula has a very simple link representation
in twistor space, where cancellations that are non-obvious in physical
space become manifest.
\end{abstract}

\maketitle

\section{Introduction}

The past several years have witnessed tremendous progress in our understanding
of the mathematical structure of scattering amplitudes, particularly in
maximally supersymmetric theories.
It is easy to argue that the seeds of this progress were
sown over two decades ago by the discovery~\cite{Parke:1986gb,Berends:1987me}
of the stunningly simple formula\footnote{Here and throughout
the paper we use calligraphic letters ${\cal A}$, ${\cal M}$ to denote
superspace amplitudes with the overall
delta-function of supermomentum conservation suppressed.}
\begin{equation}
\label{eq:PT}
{\cal A}^{\rm MHV}(1,\ldots,n)
= \frac{1}{\ket{1}{2} \ket{2}{3} \cdots \ket{n}{1}}
\end{equation}
for the maximally helicity violating (MHV)
color-ordered subamplitude for $n$-gluon scattering.
The importance of this formula goes far beyond simply knowing the answer
for a certain scattering amplitude, which one may or may not be
particularly interested in.  Rather, the mere existence of such a simple
formula for something which would normally require
enormously tedious calculations
using traditional Feynman diagram techniques suggests firstly that the
theory must possess some remarkable and deeply hidden mathematical structure,
and secondly that if one actually is interested in knowing the answer
for a certain amplitude it behooves one to discover and understand this
structure.  In other words, the formula~(\ref{eq:PT}) is as important
psychologically as it is physically, since it provides strong motivation
for digging more deeply into scattering amplitudes.

Much of the progress on gluon amplitudes can be easily recycled
and applied to graviton amplitudes due ultimately to the KLT
relations~\cite{Kawai:1985xq}
which roughly speaking state that ``gravity is Yang-Mills squared''.
Slightly more precisely, the KLT relations express an $n$-graviton amplitude
as a sum over permutations of the square of the color-ordered $n$-gluon
subamplitude times some simple extra factors (see~\cite{Bern:2002kj}
for a review).  There are several indications that maximal supergravity
may be an extraordinarily remarkable
theory~\cite{Green:2006gt,Kallosh:2007ym,Kallosh:2008ic,Naculich:2008ew,BjerrumBohr:2008ji,ArkaniHamed:2008gz,Badger:2008rn,Kallosh:2008rr,Kallosh:2008ru,Kallosh:2009db}, and possibly even ultraviolet
finite~\cite{Bern:2006kd,Green:2006yu,Green:2007zzb,Kallosh:2008mq,Bossard:2009sy,Kallosh:2009jb},
but our feeling is that even at tree level
we are still far from fully unlocking the structure of graviton
amplitudes.

To illustrate this disparity we need look no further than the simplest
graviton
amplitudes.
The original BGK (Berends, Giele and Kuijf) formula
for the $n$-graviton MHV amplitude~\cite{Berends:1988zp}
is now over 20 years old.
For later convenience we review here a
different form due to Mason and Skinner~\cite{Mason:2008jy}, who
proved the equivalence of the original BGK formula to the
expression
\begin{equation}
\label{eq:MS}
{\cal M}_n^{\rm MHV} = \sum_{P(1,\ldots,n-3)}
\frac{1}{\ket{n}{n{-}2} \ket{n{-}2}{n{-}1} \ket{n{-}1}{n}}
\frac{1}{\ket{1}{2} \cdots \ket{n}{1}}
\prod_{k=1}^{n-3}
\frac{[ k|p_{k+1} + \cdots + p_{n{-}2}|n{-}1\rangle}{\ket{k}{n{-}1}}\,,
\end{equation}
where the sum indicates a sum over all $(n-3)!$ permutations of the
labels $1,\ldots,n-3$ and we use the convention
\begin{equation}
[a|p_i+p_j+\cdots|b\rangle = \bra{a}{i}\ket{i}{b} + \bra{a}{k}\ket{j}{b}
+ \cdots.
\end{equation}

The fact that any closed form expression exists at all for this quantity,
the calculation of which would otherwise be vastly more complicated even
than the corresponding one for $n$ gluons, is an amazing achievement.
Nevertheless the formula has some features which strongly suggest that
it is not the end of the story.

First of all, the formula~(\ref{eq:MS}) does not manifest the requisite
permutation symmetry of an $n$-graviton
superamplitude.  Specifically, any superamplitude
${\cal M}_n$ must be fully symmetric under all
$n!$ permutations of the labels $1,\ldots,n$ of the external
particles, but only an $S_{n-3}$ subgroup of this symmetry is manifest
in~(\ref{eq:MS}) (several formulas which manifest a slightly larger
$S_{n-2}$ subgroup are known~\cite{Bedford:2005yy,
Elvang:2007sg}).  Of course one can check, numerically if
necessary, that~(\ref{eq:MS}) does in fact have this symmetry,
but it is far from obvious.
Moreover, even the $S_{n-3}$ symmetry arises in a somewhat contrived
way, via an explicit sum over permutations.  Undoubtedly the summand
in~(\ref{eq:MS}) contains redundant information which is washed out
by taking the sum.
This situation should be contrasted with that of Yang-Mills theory,
where~(\ref{eq:PT}) is manifestly invariant under the appropriate
dihedral symmetry
group (not the full permutation group, due to the color ordering of gluons).

Secondly,
one slightly disappointing feature of all
previously known MHV formulas including~(\ref{eq:MS})
is the appearance of ``$\cdots$'', which indicates that a particular
cyclic ordering
of the particles must be chosen in order to write the formula, even though
a graviton amplitude ultimately cannot depend on any such ordering since
gravitons do not carry any color labels.
This vestigial feature usually traces back to the use of the
KLT relations to calculate graviton amplitudes by recycling
gluon amplitudes.

An important feature
of graviton amplitudes is that they fall off like $1/z^2$ as the
supermomenta of any two particles are taking to infinity in a particular
complex direction
(see~\cite{Bedford:2005yy,Cachazo:2005ca,BjerrumBohr:2005jr,Benincasa:2007qj,Bern:2007xj,ArkaniHamed:2008yf,Bianchi:2008pu}
and~\cite{ArkaniHamed:2008gz} for the most complete
treatment),
unlike in Yang-Mills theory where the falloff is only
$1/z$~\cite{Britto:2005fq}.
It has been argued~\cite{Bern:2007xj}
that this exceptionally soft behavior of graviton
tree amplitudes is of direct importance
for the remarkable ultraviolet
cancellations in supergravity loop amplitudes~\cite{Bern:2006kd,
Bern:2007hh,Bern:2008pv,Bern:2009kf,Bern:2009kd}.

The $1/z^2$ falloff of~(\ref{eq:MS}) is manifest for each term separately
inside the sum over permutations.
Two classes of previously known formulas
for the $n$-graviton MHV amplitude
are:
those like~(\ref{eq:MS}) which manifest the $1/z^2$
falloff but only $S_{n-3}$ symmetry, and others
(see for
example~\cite{Bedford:2005yy,
Elvang:2007sg})
which have a larger $S_{n-2}$ symmetry but only
manifest falloff like $1/z$.  In the latter class of formulas
the stronger $1/z^2$ behavior arises
from delicate and non-obvious cancellations between various terms in the
sum.  This is both a feature and a bug.  It is a feature because
it implies the existence of linear identities (which have been called
bonus relations in~\cite{Spradlin:2008bu})
between individual terms in the sum which
have proven useful, for example, in establishing the equality of
various previously known but not obviously equivalent
formulas~\cite{Spradlin:2008bu}.  But it is a bug because it indicates
that the $S_{n-2}$-invariant formulas contain redundant information
distributed amongst the various terms in the sum.
The bonus relations allow one to squeeze this redundant information out
of any $S_{n-2}$-invariant formula at the cost of reducing the
manifest symmetry
to $S_{n-3}$.

It is difficult to imagine that it might be possible to improve upon
the Parke-Taylor formula~(\ref{eq:PT}) for the $n$-gluon
MHV amplitude.  However, for the reasons just reviewed, we feel
that~(\ref{eq:MS}) cannot be the end of the story for gravity.
Ideally one would like to have a formula for $n$-graviton scattering
that (1) is manifestly $S_n$ symmetric without the need for introducing
an explicit sum over permutations to impose the symmetry {\it vi et armis};
(2) makes no vestigial reference to any cyclic ordering of the $n$
gravitons, and (3) manifests $1/z^2$ falloff term by term,
making it unsqueezable by the bonus relations.

In this paper we present and prove
the ``tree formula''~(\ref{eq:main}) for the MHV scattering amplitude
which addresses the second
and third points but only manifests $S_{n-2}$ symmetry.
In section 2
we introduce the tree formula
and discuss several special cases as well
as the general soft limit.
In section 3 we work out the simple link representation of the amplitude
in twistor space, from which new physical space formula follows.
Finally the proof is in section 4.

\bigskip
\noindent
{\bf Note Added}

After this paper appeared we learned of an ansatz for the
MHV graviton amplitude presented in section 6 of~\cite{Bern:1998sv}
which upon inspection is immediately seen
to share the nice features of the
tree formula.
In fact, although
terms in the two formulas are arranged in different
ways (labeled tree diagrams versus Young tableaux), it is not
difficult to check that their content is actually identical.
Interestingly the formula of~\cite{Bern:1998sv} was constructed
with the help of ``half-soft factors'' similar in idea to the
``inverse soft limits''
which appeared much more recently
in~\cite{NimasTalk1}.
Our work establishes the validity of the ansatz conjectured
in~\cite{Bern:1998sv} and demonstrates that it arises naturally
in twistor space.

\section{The MHV Tree Formula}

\subsection{Statement of the Tree Formula}

Here we introduce a formula for the $n$-graviton
MHV scattering amplitude which we call the ``tree formula'' since it consists
of a sum of terms, each of which is conveniently represented by a tree
diagram.  The tree formula manifests an
$S_{n-2}$ subgroup of the full permutation
group.
For the moment we choose to treat particles $n-1$ and $n$ as special.
With this arbitrary choice the formula is:
\begin{equation}
\label{eq:main}
{\cal M}_n^{\rm MHV} = \frac{1}{\ket{n-1}{n}^2} \sum_{\rm trees}
\left( \prod_{{\rm edges}\,ab} \frac{\bra{a}{b}}{\ket{a}{b}}\right)
\left( \prod_{{\rm vertices}\,a} \left( \ket{a}{n-1} \ket{a}{n}
\right)^{\deg(a)-2}\right).
\end{equation}
To write down an expression for the $n$-point amplitude
one draws all inequivalent connected tree graphs with vertices labeled
$1,2,\ldots,n-2$.
(It was proven by Cayley that
there are precisely $(n-2)^{n-4}$ such diagrams.)
For example, one of the 125 labeled tree graphs
contributing to the $n=7$ graviton amplitude is
\begin{center}
\begin{picture}(120,100)(0,0)
\Line(30,50)(110,50)
\Line(30,50)(1.7157,21.7157)
\Line(30,50)(1.7157,78.2843)
\GCirc(1.7157,21.7157){10}{1.}
\GCirc(1.7157,78.2843){10}{1.}
\GCirc(30,50){10}{1.}
\GCirc(70,50){10}{1.}
\GCirc(110,50){10}{1.}
\put(1.7157,21.7157){\makebox(0,0){$2$}}
\put(1.7157,78.2843){\makebox(0,0){$5$}}
\put(30,50){\makebox(0,0){$3$}}
\put(70,50){\makebox(0,0){$1$}}
\put(110,50){\makebox(0,0){$4$}}
\end{picture}
\end{center}
According to~(\ref{eq:main})
the value of a diagram is then the product of three factors:
\begin{enumerate}
\item{an overall factor of $1/\ket{n{-}1}{n}^2$,}
\item{a factor of
$\bra{a}{b}/\ket{a}{b}$ for each propagator connecting vertices $a$ and $b$,
and}
\item{a factor of
$(\ket{a}{n{-}1} \ket{a}{n})^{\deg(a)-2}$
for each vertex $a$, where $\deg(a)$ is the degree
of the vertex (the number of edges attached to it).}
\end{enumerate}

An alternate description of the formula may be given
by noting that a vertex factor of $\ket{a}{n-1} \ket{a}{n}$ may be absorbed
into each propagator connected to that vertex.
This leads to
the equivalent formula
\begin{equation}
\label{eq:alternate}
{\cal M}_n^{\rm MHV} =
\frac{1}{\ket{n{-}1}{n}^2}
\left(\prod_{a=1}^{n-2} \frac{1}{(\ket{a}{n{-}1} \ket{a}{n})^2}
\right)
\sum_{\rm trees} \prod_{{\rm edges}\,ab}
\frac{\bra{a}{b}}{\ket{a}{b}}
\ket{a}{n{-}1} \ket{b}{n{-}1} \ket{a}{n} \ket{b}{n}.
\end{equation}

\subsection{Examples}

We defer to section IV a formal proof of the tree formula as the impatient
reader may be sufficiently convinced by seeing the formula in action here for
small $n$ and by noting that it has the correct soft limits for all $n$,
as we discuss shortly.

For each of the trivial cases $n=3,4$ there is only a single tree diagram,
\begin{equation}
{\cal M}_3^{\rm MHV}
=
{\hbox{\lower 5.pt\hbox{
\begin{picture}(20,20)(0,0)
\GCirc(10,10){10}{1.}
\put(10,10){\makebox(0,0){$1$}}
\end{picture}
}}}
= \frac{1}{(\ket{1}{2} \ket{1}{3} \ket{2}{3})^2}
\end{equation}
and
\begin{equation}
{\cal M}_4^{\rm MHV} =
{\hbox{\lower 5.pt\hbox{
\begin{picture}(60,20)(0,0)
\Line(10,10)(50,10)
\GCirc(10,10){10}{1.}
\GCirc(50,10){10}{1.}
\put(10,10){\makebox(0,0){$1$}}
\put(50,10){\makebox(0,0){$2$}}
\end{picture}
}}}
=
\frac{\bra{1}{2}}{\ket{1}{2} \ket{1}{3} \ket{1}{4} \ket{2}{3} \ket{2}{4}
\ket{3}{4}^2}
\end{equation}
respectively, which immediately reproduce the correct expressions.

For $n=5$ there are three tree diagrams
\begin{equation}
\label{eq:fivepoint}
\begin{aligned}
{\hbox{\lower 5.pt\hbox{
\begin{picture}(100,20)(0,0)
\Line(10,10)(90,10)
\GCirc(10,10){10}{1.}
\GCirc(50,10){10}{1.}
\GCirc(90,10){10}{1.}
\put(10,10){\makebox(0,0){$1$}}
\put(50,10){\makebox(0,0){$2$}}
\put(90,10){\makebox(0,0){$3$}}
\end{picture}
}}}
= \frac{\bra{1}{2}\bra{2}{3}}
{\ket{1}{2}\ket{1}{4}\ket{1}{5}\ket{2}{3}\ket{3}{4}\ket{3}{5}\ket{4}{5}^2}
\\
{\hbox{\lower 5.pt\hbox{
\begin{picture}(100,20)(0,0)
\Line(10,10)(90,10)
\GCirc(10,10){10}{1.}
\GCirc(50,10){10}{1.}
\GCirc(90,10){10}{1.}
\put(10,10){\makebox(0,0){$1$}}
\put(50,10){\makebox(0,0){$3$}}
\put(90,10){\makebox(0,0){$2$}}
\end{picture}
}}}
= \frac{\bra{1}{3}\bra{2}{3}}
{\ket{1}{3}\ket{1}{4}\ket{1}{5}\ket{2}{3}\ket{2}{4}\ket{2}{5}\ket{4}{5}^2}
\\
{\hbox{\lower 5.pt\hbox{
\begin{picture}(100,20)(0,0)
\Line(10,10)(90,10)
\GCirc(10,10){10}{1.}
\GCirc(50,10){10}{1.}
\GCirc(90,10){10}{1.}
\put(10,10){\makebox(0,0){$2$}}
\put(50,10){\makebox(0,0){$1$}}
\put(90,10){\makebox(0,0){$3$}}
\end{picture}
}}}
= \frac{\bra{1}{2}\bra{1}{3}}
{\ket{1}{2}\ket{1}{3}\ket{2}{4}\ket{2}{5}\ket{3}{4}\ket{3}{5}\ket{4}{5}^2}
\end{aligned}
\end{equation}
which can easily be verified by hand to sum to the correct expression.
Agreement between the tree formula and other known formulas
such as~(\ref{eq:MS}) may be checked numerically for slightly larger
values of $n$ by assigning random values to all of the spinor helicity
variables.  A simple implementation of the tree formula in the
Mathematica symbolic computation language is presented in the appendix.

\subsection{Relation to Other Known Formulas}

The MHV tree formula is evidently quite different in form from most
other expressions in the literature.
In particular,
no reference at all is made to any particular ordering of the particles
(there is no vestigial ``$\cdots$''), and the manifest $S_{n-2}$ arises not
because of any explicit sum over $P(1,\ldots,n-2)$ but rather from the
simple fact that the collection of labeled tree diagrams has a manifest
$S_{n-2}$ symmetry.
In our view these facts serve to highlight the essential ``gravitiness'' of
the formula, in contrast to expressions such as~(\ref{eq:MS}) which are
ultimately recycled from Yang-Mills theory.

One interesting feature of the MHV tree formula is that it is, in a sense,
minimally non-holomorphic.
Graviton MHV amplitudes, unlike their Yang-Mills
counterparts,
do not depend only the holomorphic spinor helicity variables
$\lambda_i$.  The tree formula packages all of the non-holomorphicity
into the $\bra{a}{b}$ factors associated with propagators in the tree
diagrams.  Each diagram has a unique collection of propagators and a
correspondingly unique signature of $\bra{}$'s, which only involve
$n-2$ of the $n$ labels.

Like the MHV tree formula, the Mason-Skinner formula~(\ref{eq:MS})
(unlike most other formulas in the literature, including the original
BGK formula)
has non-holomorphic dependence on only $n-2$ variables.
In our labeling of~(\ref{eq:MS}) we see that
$\widetilde{\lambda}_{n-1}$ and $\widetilde{\lambda}_n$ do not appear at all.
Of course we do not mean to say that ${\cal M}$ is ``independent'' of
these two variables since there is a suppressed overall delta function
of momentum conservation $\delta^4(\sum_i \lambda_i \widetilde{\lambda}_i)$
which one could use to shuffle some $\widetilde{\lambda}$'s into others.
Rather we mean that
the tree and MS formulas have the property that all
appearance of two of the $\widetilde{\lambda}$'s has already been completely
shuffled out.

It is an illuminating
exercise to attempt a direct term-by-term comparison
of the MHV tree formula with the MS formula~(\ref{eq:MS}).
For the first non-trivial case $n=5$ the MS formula provides the
two terms
\begin{equation}
\frac{\bra{2}{3} [1|p_2 + p_3|4\rangle}
{\ket{1}{2}\ket{1}{4}\ket{1}{5}\ket{2}{3}\ket{2}{4}\ket{3}{4}\ket{3}{5}\ket{4}{5}^2}
-
\frac{\bra{1}{3} [2|p_1 + p_3|4\rangle}
{\ket{1}{2}\ket{1}{3}\ket{1}{4}\ket{2}{4}\ket{2}{5}\ket{3}{4}\ket{3}{5}\ket{4}{5}^2}.
\end{equation}
If we now expand out the bracket $[a|p_i + p_j|b\rangle = \bra{a}{i}\ket{i}{b}
+ \bra{a}{j}\ket{j}{b}$ then we find four terms:  one of them is proportional
to $\bra{1}{2}\bra{2}{3}$ and is identical to the first
line in~(\ref{eq:fivepoint}),
another proportional to $\bra{1}{2}\bra{1}{3}$ is identical to the last
line in~(\ref{eq:fivepoint}).  The remaining two terms are both proportional
to $\bra{1}{3}\bra{2}{3}$ and may be combined as
\begin{equation}
\frac{\bra{1}{3}\bra{2}{3}
\left(\ket{1}{3}\ket{2}{5} - \ket{1}{5}\ket{2}{3}\right)}
{\ket{1}{2}\ket{1}{3}\ket{1}{4}\ket{1}{5}\ket{2}{3}\ket{2}{4}\ket{2}{5}\ket{3}{5}\ket{4}{5}^2}
\end{equation}
which with the help of a Schouten identity we recognize as precisely
the second line in~(\ref{eq:fivepoint}).

We beg the reader's pardon for allowing us to indulge in one final example.
Expanding the MS formula for $n=6$ into $\ket{}{}$'s and $\bra{}{}$'s yields a
total of 36 terms.  For example there are 6 terms proportional
to the antiholomorphic structure $\bra{1}{4}\bra{2}{4}\bra{3}{4}$, totalling
\begin{multline}
\frac{\bra{1}{4}\bra{2}{4}\bra{3}{4}\ket{4}{5}}
{\ket{1}{5}\ket{2}{5}\ket{3}{5}\ket{4}{6}\ket{5}{6}^2}
\Bigg[
\frac{1}{\ket{1}{2}\ket{1}{6}\ket{2}{3}\ket{3}{4}}
+ \frac{1}{\ket{1}{2}\ket{1}{4}\ket{2}{3}\ket{3}{6}}
-\frac{1}{\ket{1}{3}\ket{1}{6}\ket{2}{3}\ket{2}{4}}
\\
-\frac{1}{\ket{1}{3}\ket{1}{4}\ket{2}{3}\ket{2}{6}}
-\frac{1}{\ket{1}{2}\ket{1}{3}\ket{2}{6}\ket{3}{4}}
-\frac{1}{\ket{1}{2}\ket{1}{3}\ket{2}{4}\ket{3}{6}}
\Bigg].
\end{multline}
After repeated use of Schouten identities this amazingly collapses to
the single term
\begin{equation}
\frac{\bra{1}{4}\bra{2}{4}\bra{3}{4}\ket{4}{5}\ket{4}{6}}
{\ket{1}{4}\ket{1}{5}\ket{1}{6}\ket{2}{4}\ket{2}{5}\ket{2}{6}\ket{3}{4}\ket{3}{5}\ket{3}{6}\ket{5}{6}^2}
=
{\hbox{\lower 25.pt\hbox{
\begin{picture}(100,70)(0,0)
\Line(00,50)(50,10)
\Line(50,50)(50,10)
\Line(90,50)(50,10)
\GCirc(10,50){10}{1.}
\GCirc(50,50){10}{1.}
\GCirc(90,50){10}{1.}
\GCirc(50,10){10}{1.}
\put(10,50){\makebox(0,0){$1$}}
\put(50,50){\makebox(0,0){$2$}}
\put(90,50){\makebox(0,0){$3$}}
\put(50,10){\makebox(0,0){$4$}}
\end{picture}
}}}
\end{equation}

We believe that these examples are representative of the general case.
Expanding out all of the brackets in the $n$-graviton MS formula
generates a total of $[(n-3)!]^2$ terms,
but there are only
$(n-2)^{n-4}$ possible distinct antiholomorphic signatures.
Collecting terms with the same signature
and repeatedly applying Schouten identities
should collapse everything into the terms generated by the MHV tree formula.
Note that this is a huge simplification:  $(n-2)^{n-4}$ is smaller
than
$[(n-3)!]^2$
by a factor that is asymptotically $n^n$.
We certainly do not have an explicit proof of this cancellation; instead
we are relying on fact that the MS formula and the tree formula
are separately proven to be correct in order to infer how the story
should go.

To conclude this discussion we should note that we
are exploring here only the structure of the various formulas,
not making any
claims about the computational complexity of the MHV tree formula as compared
to~(\ref{eq:MS}) or any other known formula.  No practical implementation of
the MS formula would proceed by first splitting all of the brackets as we
have outlined.  Indeed a naive counting of the number of terms,
$(n-3)!$ in~(\ref{eq:MS}) versus $(n-2)^{n-4}$ for the
tree formula, suggests that for computational purposes the former is almost
certainly the clear winner despite the conceptual strengths of the latter.

\subsection{Soft Limit of the Tree Formula}

Let us consider for a moment the component amplitude
\begin{equation}
\label{eq:component}
M(1^+,\ldots,(n-2)^+,(n-1)^-,n^-) = \ket{n-1}{n}^8 {\cal M}_n^{\rm MHV}
\end{equation}
with particles $n-1$ and $n$ having negative helicity.
The universal soft factor
for gravitons is~\cite{Weinberg:1965nx,Berends:1988zp}
\begin{equation}
\label{eq:soft}
\lim_{p_1 \to 0}
\frac{M(1^+,\ldots,(n-2)^+,(n-1)^-,n^-)}{M(2^+,\ldots,(n-2)^+,(n-1)^-,n^-)}
= \sum_{i=2}^{n-2} g(i^+), \quad
g(i^+) =
\frac{\ket{i}{n-1}}{\ket{1}{n-1}}
\frac{\ket{i}{n}}{\ket{1}{n}} \frac{\bra{1}{i}}{\ket{1}{i}}.
\end{equation}
It is simple to see that the MHV tree formula satisfies this property:
the tree diagrams which do not vanish in the limit $p_1 \to 0$ are those
in which vertex 1 is connected by a propagator to
a single other vertex $i$.
Such diagrams remain connected when vertex 1 is chopped off,
leaving a contribution to the $n-1$-graviton amplitude times the
indicated factor $g(i^+)$.

Thinking about this process in reverse therefore suggests a simple
interpretation of~(\ref{eq:soft})
in terms of tree diagrams---it is a sum over all possible places
$i$ where the vertex 1 may be attached
to the $n-1$-graviton amplitude.
This structure is exactly that of the ``inverse soft factors''
suggested recently in~\cite{NimasTalk1}, and we have checked
that the MHV tree formula may be built up by recursively applying the
rule proposed there.

\section{The MHV Tree Formula in Twistor Space}

Before turning to the formal proof of the tree formula in the next section,
here we work out the link representation of the
MHV graviton amplitude in twistor space, which was one of the steps which
led to the discovery of the tree formula.
Two papers~\cite{Mason:2009sa,ArkaniHamed:2009si}
have recently constructed versions of the BCF on-shell
recursion relation directly in twistor space variables.
We follow
the standard notation where $\mu$, $\widetilde{\mu}$ are respectively
Fourier transform conjugate to the spinor helicity variables
$\lambda$, $\widetilde{\lambda}$, and assemble these
together with a four-component Grassmann variable
$\eta$ and its conjugate $\widetilde{\eta}$ into the $4|8$-component
supertwistor variables
\begin{equation}
{\cal Z} = \begin{pmatrix}
\lambda \\
\mu \\
\eta
\end{pmatrix}, \qquad
{\cal W} = \begin{pmatrix}
\widetilde{\mu} \\
\widetilde{\lambda} \\
\widetilde{\eta}
\end{pmatrix}.
\end{equation}
In the approach of~\cite{ArkaniHamed:2009si},
in which variables of both chiralities ${\cal Z}$ and ${\cal W}$ are used
simultaneously,
an apparently
important role is played by the link representation
which expresses an amplitude ${\cal M}$ in the form
\begin{equation}
{\cal M}({\cal Z}_i, {\cal W}_J) =
\int dc \ U(c_{iJ}, \lambda_i, \widetilde{\lambda}_J)
\exp\left[i  \sum_{i,J} c_{iJ} {\cal Z}_i \cdot {\cal W}_J
\right].
\end{equation}
Here one splits the $n$ particles into two groups, one of which
(labeled by $i$) one chooses to represent in ${\cal Z}$ space and the other
of which (labeled by $J$) one chooses to represent in ${\cal W}$ space.
The integral runs over all of the aptly-named link variables $c_{iJ}$ and
we refer to the integrand $U(c_{iJ}, \lambda_i, \widetilde{\lambda}_J)$
as the link representation of
${\cal M}$.
It was shown in~\cite{ArkaniHamed:2009si}
that the BCF on-shell recursion in twistor space involves nothing more than
a simple integral over ${\cal Z}$, ${\cal W}$ variables with a simple
(and essentially unique) measure factor.

The original motivation for our investigation was to explore the structure
of link representations for graviton amplitudes.
We will always adopt the
convenient convention of expressing an N${}^k$MHV amplitude in terms of
$k+2$ ${\cal Z}$ variables and $n-k-2$ ${\cal W}$ variables.
The three-particle MHV
and $\overline{\rm MHV}$ amplitudes
\begin{equation}
U_3^{\rm MHV}
= \frac{| \ket{1}{2} |}{c_{13}^2 c_{23}^2}, \qquad
U_3^{\overline{\rm MHV}}
= \frac{ | \bra{1}{2} |}{c_{31}^2 c_{32}^2}
\end{equation}
seed the on-shell recursion, which is then sufficient (in principle) to
determine the link representation for any desired amplitude.

For example, the four-particle amplitude
is the sum of two contributing BCF
diagrams
\begin{equation}
\label{eq:fourp}
U^{\rm MHV}_4=
\frac{\ket{1}{2} \bra{3}{4}}{c_{13}^2 c_{24}^2 c_{12:34}} +
\frac{\ket{1}{2} \bra{3}{4}}{c_{13}^2 c_{24}^2 c_{14} c_{23}}
\end{equation}
where we use the notation
\begin{equation}
c_{i_1 i_2:J_1 J_2} = c_{i_1 J_1} c_{i_2 J_2} - c_{i_1 J_2} c_{i_2 J_1}.
\end{equation}
Remarkably the two terms in~(\ref{eq:fourp}) combine nicely into the
simple result presented already
in~\cite{ArkaniHamed:2009si}:
\begin{equation}
U^{\rm MHV}_4
= \frac{\ket{1}{2} \bra{3}{4}}{c_{13} c_{14} c_{23} c_{24}
c_{12:34}}.
\end{equation}
This simplification seems trivial at the moment but it is just the tip
of an iceberg.  For larger $n$ the enormous simplifications discussed in
the previous section, which are apparently non-trivial in physical space,
occur automatically in the link representation.

For example the five particle MHV amplitude is
the sum of three BCF diagrams,
\begin{equation}
\begin{aligned}
U^{\rm MHV}_5=
\left\{
\frac{ | \ket{1}{2} | \bra{4}{5} ( c_{24} \bra{3}{4} + c_{25} \bra{3}{5} )}
{c_{13} c_{23} c_{14} c_{25}^2 c_{12:34} c_{12:45}}
+ (3 \leftrightarrow 4)
\right\}
+ \frac{ |\ket{1}{2} | \bra{3}{4} (c_{24} \bra{4}{5} + c_{23} \bra{3}{5})}
{c_{13} c_{14} c_{15} c_{23} c_{24} c_{25}^2 c_{12:34}}
\end{aligned}
\end{equation}
which nicely simplifies to
\begin{equation}
\frac{1}{|\ket{1}{2}|} U^{\rm MHV}_5=
\frac{ \bra{3}{4} \bra{4}{5}}{
c_{13} c_{15} c_{23} c_{25} c_{12:34} c_{12:45}}
+ \frac{ \bra{3}{5} \bra{4}{5}}{
c_{13} c_{14} c_{23} c_{24} c_{12:35} c_{12:45}}
+ \frac{ \bra{3}{4} \bra{3}{5}}{
c_{14} c_{15} c_{24} c_{25} c_{12:34} c_{12:35}}.
\end{equation}
This expression already exhibits the structure of the MHV tree formula
(except that here particles 1 and 2 are singled out, and the vertices
of the trees are labeled by $\{3,4,5\}$).

Subsequent investigations for higher $n$ reveal the general pattern
which is as follows.  Returning to the convention
where particles $n-1$ and $n$ are treated as special,
the link representation for any desired MHV amplitude may be written
down by drawing all tree diagrams with vertices labeled by
$\{1,\ldots,n-2\}$ and then assigning
\begin{enumerate}
\item{an overall factor of $\ket{n-1}{n} {\rm sign}(\ket{n-1}{n})^n$,}
\item{for each propagator connecting nodes $a$ and $b$, a factor
of $\bra{a}{b}/c_{n-1,n:a,b}$,}
\item{for each vertex $a$, a factor of $(c_{n-1,a} c_{n,a}
)^{\deg(a)-2}$,
where $\deg(a)$ is the degree of the vertex labeled $a$.
}
\end{enumerate}
It is readily verified by direct integration over the link variables that
these rules are precisely the link-space representation of the physical
space rules for the MHV tree formula given in the previous section.

\section{Proof of the MHV Tree Formula}

Here we present a proof of the MHV tree formula.
One way one might attempt to prove the formula would be to
show directly that it satisfies the BCF on-shell recursion
relation~\cite{Britto:2004ap,Britto:2005fq}
for gravity~\cite{Bedford:2005yy,Cachazo:2005ca,Benincasa:2007qj}, but
the structure of the formula is poorly suited for this task.
Instead we proceed by considering the usual BCF deformation of
the formula
${ M}^{\rm MHV}_n$ by a complex parameter $z$ and demonstrating
that ${ M}^{\rm MHV}_n(z)$ has the same residue at every pole (and
behavior at infinity) as
the similarly deformed graviton amplitude, thereby establishing equality
of the two for all $z$.

In this section we return to singling out particles 1 and 2,
letting the vertices in the tree diagrams carry the labels $\{3,\ldots,n\}$.
Then the MHV tree formula~(\ref{eq:main}) can be written as
\begin{equation}
{ M}_n^{\rm MHV} = {\ket{1}{2}^6}
\sum_{\rm trees} \frac{\bra{}{}\cdots \bra{}{}}{\ket{}{} \cdots \ket{}{}}
\prod_{a=3}^n \left(\ket{1}{a} \ket{2}{a}\right)^{\deg(a)-2}
\end{equation}
(note that we continue to work with the component
amplitude~(\ref{eq:component}))
where the factors
$\bra{}{}\cdots\bra{}{}/\ket{}{} \cdots \ket{}{}$
associated with the propagators of a
diagram are independent of 1 and 2.
Let us now make the familiar BCF shift~\cite{Britto:2005fq}
\begin{equation}
\lambda_1 \to \lambda_1(z) = \lambda_1 - z \lambda_2, \qquad
\widetilde{\lambda}_2 \to \widetilde{\lambda}_2(z) =
\widetilde{\lambda}_2 + z \widetilde{\lambda}_1
\end{equation}
which leads to the $z$-deformed MHV tree formula
\begin{equation}
\label{eq:deformedM}
{ M}_n^{\rm MHV}(z) = {\ket{1}{2}^6}
\sum_{\rm trees} \frac{\bra{}{}\cdots \bra{}{}}{\ket{}{} \cdots \ket{}{}}
\prod_{a=3}^n \left[ (\ket{1}{a} - z \ket{2}{a}) \ket{2}{a}
\right]^{\deg(a)-2}.
\end{equation}
Here we are in a position to observe a nice fact: since
each tree diagram is connected, the degrees satisfy the sum rule
\begin{equation}
\label{eq:sumrule}
\sum_{a=3}^n (\deg(a) - 2) = - 2,
\end{equation}
which guarantees that each individual term in~(\ref{eq:deformedM})
manifestly behaves like $1/z^2$ at large $z$.
This exceptionally soft behavior of graviton amplitudes
is completely hidden in the
usual Feynman diagram expansion.

\begin{figure}
\includegraphics{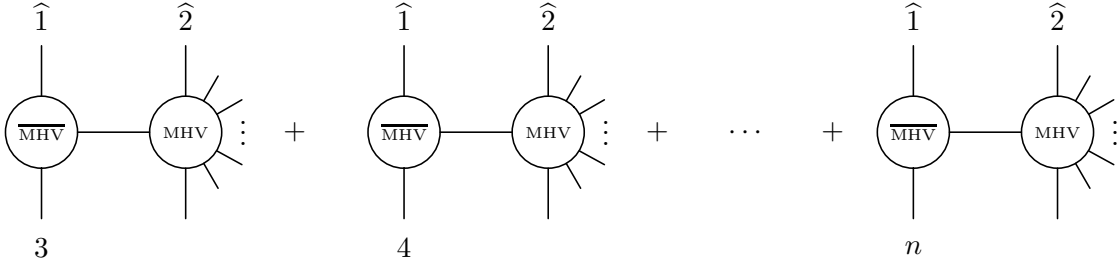}
\vskip -0.5cm
\caption{
All factorizations contributing to the on-shell recursion
relation for the $n$-point MHV amplitude.  Only the first
diagram contributes to the residue at $z = \ket{1}{3}/\ket{2}{3}$.}
\label{mhvfig2}
\end{figure}

A complex function of a single variable which vanishes at infinity is
uniquely determined by the locations of its poles as well as its residues.
Having noted that~(\ref{eq:deformedM}) has the correct behavior at large
$z$,
we can conclude the proof of the MHV tree formula by demonstrating
that~(\ref{eq:deformedM}) has precisely the
expected residues at all of its poles.
In order to say what the expected residues are we shall use induction on
$n$.  As discussed above the tree formula is readily verified for
sufficiently small $n$, so let us assume that it has been established
up through $n-1$.  We can then use BCF on-shell recursion (whose terms
are displayed graphically in~\fig{mhvfig2}) to determine what the residues
in the deformed $n$-point amplitude ought to be.

Without loss of generality let us consider just the pole at
$z = z_3 \equiv \ket{1}{3}/\ket{2}{3}$.
The only tree diagrams which contribute to the residue at this pole are those
with $\deg(3) = 1$, meaning that the vertex labeled 3 is connected to the
rest of the diagram by a single propagator.
Chopping off vertex 3 gives a subdiagram with vertices labeled
$\{4,\ldots,n\}$.  Clearly all diagrams which contribute to this residue can
be generated by first considering the collection
of tree diagrams with vertices labeled $\{4,\ldots,n\}$ and then attaching
vertex $3$ in all possible ways to the $n-3$ vertices of the subdiagram.
We therefore have
\begin{equation}
\label{eq:agree}
{ M}_n^{\rm MHV}(z) \sim {\ket{1}{2}^6}
\sum_{\rm subdiagrams} \frac{\bra{}{}\cdots \bra{}{}}{\ket{}{} \cdots \ket{}{}}
\left(\sum_{b=4}^n \frac{\bra{3}{b}}{\ket{3}{b}}
\ket{\widehat{1}}{b} \ket{2}{b} \right)
\frac{1}{\ket{\widehat{1}}{3} \ket{2}{3}}
\prod_{a=4}^n \left( \ket{\widehat{1}}{a} \ket{2}{a} \right)^{\deg(a)-2}
\end{equation}
where $\sim$ denotes that we have dropped terms which are nonsingular
at $z = z_3$, the sum over $b$ runs over all the places where vertex
3 can be attached to the subdiagram, and
$\bra{}{}\cdots\bra{}{}/\ket{}{}\cdots\ket{}{}$
indicates all edge factors associated the subdiagram, necessarily independent
of 3.
Using the Schouten identity we find that $\ket{\widehat{1}}{b} =
\ket{1}{2} \ket{b}{3}/\ket{2}{3}$ so we have
after a couple of simple steps (and using~(\ref{eq:sumrule}))
\begin{equation}
{ M}_n^{\rm MHV}(z) \sim {\ket{1}{2}^6} \frac{ \bra{1}{3}}
{\ket{1}{3} - z \ket{2}{3}}
\sum_{\rm subdiagrams}
\frac{\bra{}{}\cdots \bra{}{}}{\ket{}{} \cdots \ket{}{}}
\prod_{a=4}^n \left(\ket{2}{a} \ket{3}{a} \right)^{\deg(a)-2}.
\end{equation}

On the other hand we know from the on-shell recursion for the $n$-point
amplitude that the residue at $z=z_3$ comes entirely from the first
BCFW diagram in~\fig{mhvfig2}, whose value is
\begin{equation}
\label{eq:temp}
{ M}^{\overline{\rm MHV}}_3(z_3) \times
\frac{1}{P^2(z)} \times
{ M}^{\rm MHV}_{n-1}(z_3)
\end{equation}
where
\begin{equation}
\label{eq:phat}
P(z) = p_1 + p_3 - z \lambda_2 \widetilde{\lambda}_1.
%\widehat{P} = p_1 + p_3 - \frac{\ket{1}{3}}{\ket{2}{3}}
%\lambda_2 \widetilde{\lambda}_1.
\end{equation}
Assuming the validity of the MHV tree formula for the $n-1$-point amplitude
on the right, the expression~(\ref{eq:temp}) evaluates to
\begin{equation}
\frac{\bra{\widehat{P}}{3}^6}{ \bra{3}{1}^2 \bra{1}{\widehat{P}}^2}
\times
\frac{1}{\bra{1}{3} (\ket{1}{3} - z \ket{2}{3})}
\times
{\ket{\widehat{P}}{2}^6}
\sum_{\rm subdiagrams}
\frac{\bra{}{}\cdots \bra{}{}}{\ket{}{} \cdots \ket{}{}}
\prod_{a=4}^n \left(\ket{\widehat{P}}{a} \ket{2}{a} \right)^{\deg(a)-2}
\end{equation}
where $\widehat{P} = P(z_3)$.
After simplifying this result with the help of~(\ref{eq:phat}) we
find precise agreement with~(\ref{eq:agree}), thereby completing the proof
of the MHV tree formula.

\section{Discussion and Open Questions}

The tree formula introduced in this paper has several
conceptually satisfying features and almost completely
fulfills the wish-list outlined in the introduction.
It appears to be a genuinely gravitational
formula, rather than a recycled Yang-Mills result.
Is it, finally, the end of the story for the the
MHV amplitude, as the Parke-Taylor
formula~(\ref{eq:PT})
surely is for the $n$-gluon MHV amplitude?

Among the wish-list items
the MHV tree formula fails only in manifesting the full $S_n$ symmetry.
Of course it is possible
that there simply does not exist any
natural more primitive formula which manifests the full symmetry.
It is not obvious how one could go about constructing such a formula, but
we can draw some encouragement
and inspiration from the recent paper~\cite{Hodges:2009hk}
which demonstrates how to write manifestly dihedral symmetric
formulas for NMHV amplitudes in Yang-Mills theory as certain volume
integrals in twistor space.
Different ways of dividing the volume
into tetrahedra give rise to
apparently different but equivalent formulas for NMHV amplitudes.
The same goal can apparently also be achieved by writing the amplitude
as a certain contour integral where different choices of contour produce
different looking but actually
equivalent formulas~\cite{NimasTalk2,NimasTalk3}.
Perhaps in gravity even the MHV amplitude needs to
be formulated in a way which is fundamentally
symmetric but which nevertheless requires choosing two of the
$n$ gravitons for special treatment.

In Yang-Mills theory the only formula we know of which manifests the
full dihedral symmetry for all superamplitudes is the connected
prescription~\cite{Roiban:2004vt,Roiban:2004ka,Roiban:2004yf,Roiban:2004jh,Spradlin:2005hi}
which follows from Witten's formulation of Yang-Mills theory
as a twistor string theory~\cite{Witten:2003nn}.
Perhaps finding fully $S_n$ symmetric formulas for graviton superamplitudes
requires the construction of an appropriate
twistor string theory for supergravity,
an important question in its own right which has attracted some
attention~\cite{Giombi:2004ix,BjerrumBohr:2006sg,AbouZeid:2006wu,Nair:2007md,Mason:2008jy}.
An important motivation for Witten's twistor string theory was provided
by Nair's observation~\cite{Nair:1988bq}
that the Parke-Taylor formula~(\ref{eq:PT}) could be computed
as a current algebra correlator in a WZW model.
The BGK formula (essentially~(\ref{eq:MS})) can similarly be related
to current correlators and vertex operators in twistor
space~\cite{Nair:2005iv}, but
we hope that the new MHV tree formula might provide a more
appropriate starting point for
this purpose and perhaps shed some more light on a
twistor-string-like description for supergravity.

Another obvious avenue for future research is to investigate whether
any of the advances made here can be usefully applied to non-MHV
amplitudes.  Unfortunately we have not yet found any very nice structure
in the link representation for non-MHV graviton amplitudes.
Recently in~\cite{Drummond:2009ge} it was demonstrated how to solve
the on-shell recursion for
all tree-level supergraviton amplitudes, following steps very similar
to those which were used to solve the recursion for supersymmetric
Yang-Mills~\cite{Drummond:2008cr}.
In~\cite{Drummond:2009ge} a crucial role was played by what was called
the graviton subamplitude, which is the summand of an $n$-particle
graviton amplitude inside a sum over $(n-2)!$ permutations.
The decomposition of every amplitude into its subamplitudes
allowed for a very efficient application of the on-shell recursion since the
same two legs could be singled out and shifted at each step in the recursion.
Unfortunately there is no natural notion of a subamplitude for the
MHV tree formula, making it very poorly suited as a starting point
for attempting to solve the on-shell recursion.
In our view the fact that the tree formula apparently can neither be
easily derived from BCF, nor usefully used as an input to
BCF, suggests the possible
existence of some kind of new rules for the efficient calculation of more
general gravity amplitudes.

The arrangement of supergravity amplitudes into ordered subamplitudes
also proved very useful in~\cite{Hall:2009xg,Katsaroumpas:2009iy}
for the purpose of
expressing the coefficients of one-loop supergravity amplitudes in terms
of one-loop Yang-Mills coefficients.  It would certainly be very interesting
to see if any of aspects of the MHV tree formula could be useful for
loop amplitudes in supergravity, if at least as input for
unitarity sums~\cite{Elvang:2008na,Bern:2009xq}.

\section*{Acknowledgments}

We are grateful to Nima Arkani-Hamed, Freddy Cachazo, Clifford Cheung
and Cristian Vergu
for numerous inspiring and explanatory conversations, and
to Zvi Bern and Lance Dixon for bringing~\cite{Bern:1998sv}
to our attention.
This work was supported in part by the US
Department of Energy under contract DE-FG02-91ER40688 (MS (OJI)
and AV), and the US National Science Foundation under grants
PHY-0638520 (MS) and PHY-0643150 CAREER and PECASE (AV).

\appendix

\section*{The MHV Tree Formula in Mathematica}

Here we present for the reader's benefit a simple command
implementing the MHV tree formula in Mathematica:
\begin{verbatim}
Needs["Combinatorica`"];
MHV[n_Integer]/;n>4 := 1/ket[n-1,n]^2 1/(Times @@ ((ket[n-1,#] ket[n,#])^2
     & /@ Range[n-2])) ((Times @@ (Transpose[#]/.{a___,1,b___,1,c___} :>
     prop[Length[{a}]+1,Length[{a,b}]+2])) & /@ IncidenceMatrix /@
     CodeToLabeledTree /@ Flatten[Outer[List,Sequence @@
     Table[Range[n-2],{n-4}]],n-5]) /. prop[a_,b_] ->
     bra[a,b]/ket[a,b] ket[n-1,a] ket[n-1,b] ket[n,a] ket[n,b];
\end{verbatim}
Here we use the notation ${\tt ket[}a,b{\tt ]} = \ket{a}{b}$
and ${\tt bra[}a,b{\tt ]} = \bra{a}{b}$.
The (trivial) cases $n=3,4$ must be handled separately.


\begin{thebibliography}{99}

\bibitem{Parke:1986gb}
  S.~J.~Parke and T.~R.~Taylor,
  %``An Amplitude for $n$ Gluon Scattering,''
  Phys.\ Rev.\ Lett.\  {\bf 56}, 2459 (1986).
  %%CITATION = PRLTA,56,2459;%%

\bibitem{Berends:1987me}
  F.~A.~Berends and W.~T.~Giele,
  %``Recursive Calculations for Processes with n Gluons,''
  Nucl.\ Phys.\  B {\bf 306}, 759 (1988).
  %%CITATION = NUPHA,B306,759;%%

\bibitem{Kawai:1985xq}
  H.~Kawai, D.~C.~Lewellen and S.~H.~H.~Tye,
  %``A Relation Between Tree Amplitudes Of Closed And Open Strings,''
  Nucl.\ Phys.\  B {\bf 269}, 1 (1986).
  %%CITATION = NUPHA,B269,1;%%

\bibitem{Bern:2002kj}
  Z.~Bern,
  %``Perturbative quantum gravity and its relation to gauge theory,''
  Living Rev.\ Rel.\  {\bf 5}, 5 (2002)
  [arXiv:gr-qc/0206071].
  %%CITATION = 00222,5,5;%%

\bibitem{Green:2006gt}
  M.~B.~Green, J.~G.~Russo and P.~Vanhove,
  %``Non-renormalisation conditions in type II string theory and maximal
  %supergravity,''
  JHEP {\bf 0702}, 099 (2007)
  [arXiv:hep-th/0610299].
  %%CITATION = JHEPA,0702,099;%%

\bibitem{Kallosh:2007ym}
  R.~Kallosh,
  %``The Effective Action of N=8 Supergravity,''
  arXiv:0711.2108.
  %%CITATION = ARXIV:0711.2108;%%

\bibitem{Kallosh:2008ic}
  R.~Kallosh and M.~Soroush,
  %``Explicit Action of E7(7) on N=8 Supergravity Fields,''
  Nucl.\ Phys.\  B {\bf 801}, 25 (2008)
  [arXiv:0802.4106].
  %%CITATION = NUPHA,B801,25;%%

\bibitem{Naculich:2008ew}
  S.~G.~Naculich, H.~Nastase and H.~J.~Schnitzer,
  %``Two-loop graviton scattering relation and IR behavior in N=8
  %supergravity,''
  Nucl.\ Phys.\  B {\bf 805}, 40 (2008)
  [arXiv:0805.2347].
  %%CITATION = NUPHA,B805,40;%%

\bibitem{BjerrumBohr:2008ji}
  N.~E.~J.~Bjerrum-Bohr and P.~Vanhove,
  %``Absence of Triangles in Maximal Supergravity Amplitudes,''
  JHEP {\bf 0810}, 006 (2008)
  [arXiv:0805.3682].
  %%CITATION = JHEPA,0810,006;%%

\bibitem{ArkaniHamed:2008gz}
  N.~Arkani-Hamed, F.~Cachazo and J.~Kaplan,
  %``What is the Simplest Quantum Field Theory?,''
  arXiv:0808.1446.
  %%CITATION = ARXIV:0808.1446;%%

\bibitem{Badger:2008rn}
  S.~Badger, N.~E.~J.~Bjerrum-Bohr and P.~Vanhove,
  %``Simplicity in the Structure of QED and Gravity Amplitudes,''
  arXiv:0811.3405.
  %%CITATION = ARXIV:0811.3405;%%

\bibitem{Kallosh:2008rr}
  R.~Kallosh and T.~Kugo,
  %``The footprint of E7 in amplitudes of N=8 supergravity,''
  JHEP {\bf 0901}, 072 (2009)
  [arXiv:0811.3414].
  %%CITATION = JHEPA,0901,072;%%

\bibitem{Kallosh:2008ru}
  R.~Kallosh, C.~H.~Lee and T.~Rube,
  %``N=8 Supergravity 4-point Amplitudes,''
  JHEP {\bf 0902}, 050 (2009)
  [arXiv:0811.3417].
  %%CITATION = JHEPA,0902,050;%%

\bibitem{Kallosh:2009db}
  R.~Kallosh,
  %``N=8 Supergravity on the Light Cone,''
  arXiv:0903.4630.
  %%CITATION = ARXIV:0903.4630;%%

\bibitem{Bern:2006kd}
  Z.~Bern, L.~J.~Dixon and R.~Roiban,
  %``Is N = 8 Supergravity Ultraviolet Finite?,''
  Phys.\ Lett.\  B {\bf 644}, 265 (2007)
  [arXiv:hep-th/0611086].
  %%CITATION = PHLTA,B644,265;%%

\bibitem{Green:2006yu}
  M.~B.~Green, J.~G.~Russo and P.~Vanhove,
  %``Ultraviolet properties of maximal supergravity,''
  Phys.\ Rev.\ Lett.\  {\bf 98}, 131602 (2007)
  [arXiv:hep-th/0611273].
  %%CITATION = PRLTA,98,131602;%%

\bibitem{Green:2007zzb}
  M.~B.~Green, H.~Ooguri and J.~H.~Schwarz,
  %``Decoupling Supergravity from the Superstring,''
  Phys.\ Rev.\ Lett.\  {\bf 99}, 041601 (2007)
  [arXiv:0704.0777].
  %%CITATION = PRLTA,99,041601;%%

\bibitem{Kallosh:2008mq}
  R.~Kallosh,
  %``On a possibility of a UV finite N=8 supergravity,''
  arXiv:0808.2310.
  %%CITATION = ARXIV:0808.2310;%%

\bibitem{Bossard:2009sy}
  G.~Bossard, P.~S.~Howe and K.~S.~Stelle,
  %``The ultra-violet question in maximally supersymmetric field theories,''
  Gen.\ Rel.\ Grav.\  {\bf 41}, 919 (2009)
  [arXiv:0901.4661].
  %%CITATION = GRGVA,41,919;%%

\bibitem{Kallosh:2009jb}
  R.~Kallosh,
  %``On UV Finiteness of the Four Loop N=8 Supergravity,''
  arXiv:0906.3495.
  %%CITATION = ARXIV:0906.3495;%%

\bibitem{Berends:1988zp}
  F.~A.~Berends, W.~T.~Giele and H.~Kuijf,
  %``On relations between multi - gluon and multigraviton scattering,''
  Phys.\ Lett.\  B {\bf 211}, 91 (1988).
  %%CITATION = PHLTA,B211,91;%%

\bibitem{Mason:2008jy}
  L.~Mason, D.~Skinner,
  %``Gravity, Twistors and the MHV Formalism,''
  arXiv:0808.3907.
  %%CITATION = ARXIV:0808.3907;%%

\bibitem{Bedford:2005yy}
  J.~Bedford, A.~Brandhuber, B.~J.~Spence and G.~Travaglini,
  %``A recursion relation for gravity amplitudes,''
  Nucl.\ Phys.\  B {\bf 721}, 98 (2005)
  [arXiv:hep-th/0502146].
  %%CITATION = NUPHA,B721,98;%%

\bibitem{Elvang:2007sg}
  H.~Elvang and D.~Z.~Freedman,
  %``Note on graviton MHV amplitudes,''
  JHEP {\bf 0805}, 096 (2008)
  [arXiv:0710.1270].
  %%CITATION = JHEPA,0805,096;%%

\bibitem{Cachazo:2005ca}
  F.~Cachazo and P.~Svrcek,
  %``Tree level recursion relations in general relativity,''
  arXiv:hep-th/0502160.
  %%CITATION = HEP-TH/0502160;%%

\bibitem{BjerrumBohr:2005jr}
  N.~E.~J.~Bjerrum-Bohr, D.~C.~Dunbar, H.~Ita, W.~B.~Perkins and K.~Risager,
  %``MHV-vertices for gravity amplitudes,''
  JHEP {\bf 0601}, 009 (2006)
  [arXiv:hep-th/0509016].
  %%CITATION = JHEPA,0601,009;%%

\bibitem{Benincasa:2007qj}
  P.~Benincasa, C.~Boucher-Veronneau and F.~Cachazo,
  %``Taming tree amplitudes in general relativity,''
  JHEP {\bf 0711}, 057 (2007)
  [arXiv:hep-th/0702032].
  %%CITATION = JHEPA,0711,057;%%

\bibitem{Bern:2007xj}
  Z.~Bern, J.~J.~Carrasco, D.~Forde, H.~Ita and H.~Johansson,
  %``Unexpected Cancellations in Gravity Theories,''
  Phys.\ Rev.\  D {\bf 77}, 025010 (2008)
  [arXiv:0707.1035].

\bibitem{ArkaniHamed:2008yf}
  N.~Arkani-Hamed and J.~Kaplan,
  %``On Tree Amplitudes in Gauge Theory and Gravity,''
  JHEP {\bf 0804}, 076 (2008)
  [arXiv:0801.2385].
  %%CITATION = JHEPA,0804,076;%%

\bibitem{Bianchi:2008pu}
  M.~Bianchi, H.~Elvang and D.~Z.~Freedman,
  %``Generating Tree Amplitudes in N=4 SYM and N = 8 SG,''
  JHEP {\bf 0809}, 063 (2008)
  [arXiv:0805.0757].
  %%CITATION = JHEPA,0809,063;%%

\bibitem{Britto:2005fq}
  R.~Britto, F.~Cachazo, B.~Feng and E.~Witten,
  %``Direct proof of tree-level recursion relation in Yang-Mills theory,''
  Phys.\ Rev.\ Lett.\  {\bf 94}, 181602 (2005)
  [arXiv:hep-th/0501052].
  %%CITATION = PRLTA,94,181602;%%

\bibitem{Bern:2007hh}
  Z.~Bern, J.~J.~Carrasco, L.~J.~Dixon, H.~Johansson, D.~A.~Kosower and R.~Roiban,
  %``Three-Loop Superfiniteness of N=8 Supergravity,''
  Phys.\ Rev.\ Lett.\  {\bf 98}, 161303 (2007)
  [arXiv:hep-th/0702112].
  %%CITATION = PRLTA,98,161303;%%

\bibitem{Bern:2008pv}
  Z.~Bern, J.~J.~M.~Carrasco, L.~J.~Dixon, H.~Johansson and R.~Roiban,
  %``Manifest Ultraviolet Behavior for the Three-Loop Four-Point Amplitude of
  %N=8 Supergravity,''
  Phys.\ Rev.\  D {\bf 78}, 105019 (2008)
  [arXiv:0808.4112].
  %%CITATION = PHRVA,D78,105019;%%

\bibitem{Bern:2009kf}
  Z.~Bern, J.~J.~M.~Carrasco and H.~Johansson,
  %``Progress on Ultraviolet Finiteness of Supergravity,''
  arXiv:0902.3765.
  %%CITATION = ARXIV:0902.3765;%%

\bibitem{Bern:2009kd}
  Z.~Bern, J.~J.~Carrasco, L.~J.~Dixon, H.~Johansson and R.~Roiban,
  %``The Ultraviolet Behavior of N=8 Supergravity at Four Loops,''
  arXiv:0905.2326.
  %%CITATION = ARXIV:0905.2326;%%

\bibitem{Spradlin:2008bu}
  M.~Spradlin, A.~Volovich and C.~Wen,
  %``Three Applications of a Bonus Relation for Gravity Amplitudes,''
  Phys.\ Lett.\  B {\bf 674}, 69 (2009)
  [arXiv:0812.4767].
  %%CITATION = PHLTA,B674,69;%%

\bibitem{Weinberg:1965nx}
  S.~Weinberg,
  %``Infrared photons and gravitons,''
  Phys.\ Rev.\  {\bf 140}, B516 (1965).
  %%CITATION = PHRVA,140,B516;%%

\bibitem{NimasTalk1}
  N.~Arkani-Hamed,
  {\tt http://www.ippp.dur.ac.uk/Workshops/09/Amplitudes/},
  talk at {\it Amplitudes 09}.

\bibitem{Mason:2009sa}
  L.~Mason and D.~Skinner,
  %``Scattering Amplitudes and BCFW Recursion in Twistor Space,''
  arXiv:0903.2083.
  %%CITATION = ARXIV:0903.2083;%%

\bibitem{ArkaniHamed:2009si}
  N.~Arkani-Hamed, F.~Cachazo, C.~Cheung and J.~Kaplan,
  %``The S-Matrix in Twistor Space,''
  arXiv:0903.2110.
  %%CITATION = ARXIV:0903.2110;%%

\bibitem{Hodges:2005bf}
  A.~P.~Hodges,
  %``Twistor diagram recursion for all gauge-theoretic tree amplitudes,''
  arXiv:hep-th/0503060.
  %%CITATION = HEP-TH/0503060;%%

\bibitem{Hodges:2005aj}
  A.~P.~Hodges,
  %``Twistor diagrams for all tree amplitudes in gauge theory: A
  %helicity-independent formalism,''
  arXiv:hep-th/0512336.
  %%CITATION = HEP-TH/0512336;%%

\bibitem{Hodges:2006tw}
  A.~P.~Hodges,
  %``Scattering amplitudes for eight gauge fields,''
  arXiv:hep-th/0603101.
  %%CITATION = HEP-TH/0603101;%%

\bibitem{Britto:2004ap}
  R.~Britto, F.~Cachazo and B.~Feng,
  %``New recursion relations for tree amplitudes of gluons,''
  Nucl.\ Phys.\  B {\bf 715}, 499 (2005)
  [arXiv:hep-th/0412308].
  %%CITATION = NUPHA,B715,499;%%

\bibitem{Hodges:2009hk}
  A.~Hodges,
  %``Eliminating spurious poles from gauge-theoretic amplitudes,''
  arXiv:0905.1473.
  %%CITATION = ARXIV:0905.1473;%%

\bibitem{NimasTalk2}
  N.~Arkani-Hamed,
  {\tt http://strings2009.roma2.infn.it/},
  talk at {\it Strings 09}.

\bibitem{NimasTalk3}
  N.~Arkani-Hamed,
  {\tt http://int09.aei.mpg.de/},
  talk at {\it Integrability in Gauge and String Theory}.

\bibitem{Roiban:2004vt}
  R.~Roiban, M.~Spradlin and A.~Volovich,
  %``A googly amplitude from the B-model in twistor space,''
  JHEP {\bf 0404}, 012 (2004)
  [arXiv:hep-th/0402016].
  %%CITATION = JHEPA,0404,012;%%

\bibitem{Roiban:2004ka}
  R.~Roiban and A.~Volovich,
  %``All googly amplitudes from the B-model in twistor space,''
  Phys.\ Rev.\ Lett.\  {\bf 93}, 131602 (2004)
  [arXiv:hep-th/0402121].
  %%CITATION = PRLTA,93,131602;%%

\bibitem{Roiban:2004yf}
  R.~Roiban, M.~Spradlin and A.~Volovich,
  %``On the tree-level S-matrix of Yang-Mills theory,''
  Phys.\ Rev.\  D {\bf 70}, 026009 (2004)
  [arXiv:hep-th/0403190].
  %%CITATION = PHRVA,D70,026009;%%

\bibitem{Roiban:2004jh}
  R.~Roiban, M.~Spradlin and A.~Volovich,
  %``Yang-Mills amplitudes from twistor string theory,''
%\href{http://www.slac.stanford.edu/spires/find/hep/www?irn=7138881}{SPIRES entry}
{\it Prepared for AMS - IMS - SIAM Summer Research Conference on String Geometry, Snowbird, Utah, 5-11 Jun 2004}

\bibitem{Spradlin:2005hi}
  M.~Spradlin,
  %``Yang-Mills amplitudes from string theory in twistor space,''
  Int.\ J.\ Mod.\ Phys.\  A {\bf 20}, 3416 (2005).
  %%CITATION = IMPAE,A20,3416;%%

\bibitem{Witten:2003nn}
  E.~Witten,
  %``Perturbative gauge theory as a string theory in twistor space,''
  Commun.\ Math.\ Phys.\  {\bf 252}, 189 (2004)
  [arXiv:hep-th/0312171].
  %%CITATION = CMPHA,252,189;%%

\bibitem{Giombi:2004ix}
  S.~Giombi, R.~Ricci, D.~Robles-Llana and D.~Trancanelli,
  %``A note on twistor gravity amplitudes,''
  JHEP {\bf 0407}, 059 (2004)
  [arXiv:hep-th/0405086].
  %%CITATION = JHEPA,0407,059;%%

\bibitem{BjerrumBohr:2006sg}
  N.~E.~J.~Bjerrum-Bohr, D.~C.~Dunbar and H.~Ita,
  %``Perturbative gravity and twistor space,''
  Nucl.\ Phys.\ Proc.\ Suppl.\  {\bf 160}, 215 (2006)
  [arXiv:hep-th/0606268].
  %%CITATION = NUPHZ,160,215;%%

\bibitem{AbouZeid:2006wu}
  M.~Abou-Zeid, C.~M.~Hull and L.~J.~Mason,
  %``Einstein supergravity and new twistor string theories,''
  Commun.\ Math.\ Phys.\  {\bf 282}, 519 (2008)
  [arXiv:hep-th/0606272].
  %%CITATION = CMPHA,282,519;%%

\bibitem{Nair:2007md}
  V.~P.~Nair,
  %``A Note on Graviton Amplitudes for New Twistor String Theories,''
  Phys.\ Rev.\  D {\bf 78}, 041501 (2008)
  [arXiv:0710.4961].
  %%CITATION = PHRVA,D78,041501;%%

\bibitem{Nair:1988bq}
  V.~P.~Nair,
  %``A CURRENT ALGEBRA FOR SOME GAUGE THEORY AMPLITUDES,''
  Phys.\ Lett.\  B {\bf 214}, 215 (1988).
  %%CITATION = PHLTA,B214,215;%%

\bibitem{Nair:2005iv}
  V.~P.~Nair,
  %``A note on MHV amplitudes for gravitons,''
  Phys.\ Rev.\  D {\bf 71}, 121701 (2005)
  [arXiv:hep-th/0501143].
  %%CITATION = PHRVA,D71,121701;%%

\bibitem{Drummond:2008cr}
  J.~M.~Drummond and J.~M.~Henn,
  %``All tree-level amplitudes in N=4 SYM,''
  JHEP {\bf 0904}, 018 (2009)
  [arXiv:0808.2475].
  %%CITATION = JHEPA,0904,018;%%

\bibitem{Drummond:2009ge}
  J.~M.~Drummond, M.~Spradlin, A.~Volovich and C.~Wen,
  %``Tree-Level Amplitudes in N=8 Supergravity,''
  arXiv:0901.2363.
  %%CITATION = ARXIV:0901.2363;%%

\bibitem{Hall:2009xg}
  A.~Hall,
  %``Supersymmetric Yang-Mills and Supergravity Amplitudes at One Loop,''
  arXiv:0906.0204.
  %%CITATION = ARXIV:0906.0204;%%

\bibitem{Katsaroumpas:2009iy}
  P.~Katsaroumpas, B.~Spence and G.~Travaglini,
  %``One-loop N=8 supergravity coefficients from N=4 super Yang-Mills,''
  arXiv:0906.0521.
  %%CITATION = ARXIV:0906.0521;%%

\bibitem{Elvang:2008na}
  H.~Elvang, D.~Z.~Freedman and M.~Kiermaier,
  %``Recursion Relations, Generating Functions, and Unitarity Sums in N=4 SYM
  %Theory,''
  JHEP {\bf 0904}, 009 (2009)
  [arXiv:0808.1720].
  %%CITATION = JHEPA,0904,009;%%

\bibitem{Bern:2009xq}
  Z.~Bern, J.~J.~M.~Carrasco, H.~Ita, H.~Johansson and R.~Roiban,
  %``On the Structure of Supersymmetric Sums in Multi-Loop Unitarity Cuts,''
  arXiv:0903.5348.
  %%CITATION = ARXIV:0903.5348;%%

\bibitem{Bern:1998sv}
  Z.~Bern, L.~J.~Dixon, M.~Perelstein and J.~S.~Rozowsky,
  %``Multi-leg one-loop gravity amplitudes from gauge theory,''
  Nucl.\ Phys.\  B {\bf 546}, 423 (1999)
  [arXiv:hep-th/9811140].
  %%CITATION = NUPHA,B546,423;%%

\end{thebibliography}
\end{document}